\documentclass[prl,twocolumn,amsmath]{revtex4}
\usepackage{graphicx}

\begin{document}

\title{\large\bf Geometric Quantum Discord in Schwarzschild de Sitter Spacetime}

\author{Xukun Wang$^{1}$, Tinggui Zhang$^{1}$ and Xiaofen Huang$^{1,\dag}$}
\affiliation{${}^{1}$ School of Mathematics and Statistics, Hainan Normal University, Haikou, 571158, China\\
$^{\dag}$ Correspondence to hangxf1206@163.com}


\begin{abstract}
We investigate geometric quantum discord for massless Dirac field modes in Schwarzschild–de Sitter spacetime, focusing on the distinct thermal effects of the black hole and cosmological horizons. For pure Bell-type and isotropic initial states, increasing the black hole mass at fixed cosmological constant restores accessible discord, while higher dimensionless frequencies favor its preservation. Inaccessible discord depends on the retained modes: higher frequencies suppress discord involving the inaccessible cosmological mode, whereas the frequency ordering reverses with increasing black hole mass for the inaccessible black hole mode paired with the accessible cosmological mode. For the selected parameters yielding equal initial discord, the two state families exhibit identical discord in sectors containing the inaccessible black hole mode but differ over part of the parameter range in the remaining inaccessible sector. At fixed positive dimensionless frequency, accessible discord returns to its initial value in the Nariai limit, while all three inaccessible discords vanish. These results reveal the interplay between horizon thermal effects, mode selection, and initial-state structure in shaping quantum correlations.
\end{abstract}

\maketitle
\section{I. Introduction}
Quantum discord captures quantum correlations beyond entanglement through the disturbance associated with local measurements \cite{OllivierZurek2001,HendersonVedral2001}. Its connections to information processing and quantum estimation have made it a useful tool for studying mixed states \cite{Modi2012,Adesso2016,Datta2008,Dakic2012,Girolami2014}. Geometric quantum discord quantifies the distance to states with vanishing quantum discord and admits a closed expression for two qubits in terms of their Bloch coefficients \cite{Dakic2010,LuoFu2010}. Extensions to multipartite systems further connect measurement sequences with the structure and dynamics of quantum correlations \cite{Radhakrishnan2020,LiFei2021}.

In relativistic settings, horizons restrict access to field modes and introduce thermal effects through the Unruh and Hawking mechanisms \cite{Unruh1976,Hawking1975,Crispino2008}. Tracing over inaccessible modes changes the correlations available to observers, with the response depending on field statistics, mode selection and localization \cite{Fuentes2005,Alsing2006,Martin2010,Bruschi2010}. Studies of quantum discord in noninertial frames and black hole backgrounds have shown that different correlation measures can respond differently to the same horizon transformation \cite{Datta2009,Brown2012,WangDiscord2010,Wang2014}. Recent work has extended these analyses to environmental decoherence, noisy mixed states and the black hole quantum atmosphere \cite{Haddadi2024,Xiong2026,LiHuang2026}.

The role of inaccessible modes and the choice of initial state are central to understanding quantum correlations in a gravitational background. Studies of entanglement, coherence and steering have found distinct responses in accessible and inaccessible modes \cite{Wang2010,Harikrishnan2022,WuSteering2022,WuAccessible2024}. The initial state also affects the response to horizon thermal effects: Hawking radiation can enhance teleportation fidelity for suitable initial states \cite{WuTeleportation2023}, and nonmaximally entangled Bell type states can retain more entanglement or steering than maximally entangled states in certain parameter ranges \cite{WuLi2025,LiShangWu2025}. Related detector studies have found entanglement enhancement or preservation for particular initial states under acceleration \cite{LiAcceleration2025,LiFreeze2026}. These results motivate comparisons of different initial states at equal initial geometric quantum discord, so that differences in their response can be examined at a common initial value of the measure.

Cosmological horizons provide another source of thermal particle creation \cite{GibbonsHawking1977}, with consequences for fermionic entanglement, coherence and multipartite correlations in de Sitter spacetime \cite{BhattacharyaChakrabortty2020,BhattacharyaGaur2020,WuCoherence2023,Jiang2025}. Schwarzschild de Sitter (SdS) spacetime contains both a black hole event horizon (BEH) and a cosmological event horizon (CEH), whose generally unequal surface gravities introduce two thermal scales \cite{Bhattacharya2018,QiuTraschen2020}. The Nariai parameter $3M\sqrt{\Lambda}$ characterizes the approach to coincident horizons. Research on scalar and Dirac fields in this geometry has examined entanglement and mutual information \cite{BhattacharyaJoshi2022,Liu2023}, followed by Gaussian and fermionic steering and genuine multipartite Bell nonlocality \cite{Wu2024,WuFermionic2025,refmi2025}. These studies motivate two questions for geometric quantum discord: how do the unequal thermal effects of the horizons influence its behavior in accessible and inaccessible modes, and can different initial states with equal initial geometric quantum discord respond differently to these effects?

We study the effects of the black hole and cosmological horizons on geometric quantum discord for massless Dirac field modes initially prepared in either pure Bell type states or isotropic states. The Bell type pure states and isotropic states allow us to compare pure and mixed initial states and to select states with equal initial geometric quantum discord. This comparison probes the influence of the initial state beyond the initial value of geometric quantum discord. Within the single mode approximation and independent treatment of the two horizon neighborhoods, we construct the four mode state $\rho_{ALBR}$ and trace over the unobserved modes. Here $A$ and $B$ are accessible modes near the black hole and cosmological horizons, respectively. Their inaccessible partners are $L$ inside the black hole and $R$ beyond the cosmological horizon. We evaluate the geometric quantum discord of the reduced states using their Bloch representation and examine how the Nariai parameter $3M\sqrt{\Lambda}$, dimensionless frequency $\omega/\sqrt{\Lambda}$ and initial state parameters $p$ and $q$ jointly determine the accessible and inaccessible correlations.

For the initial states considered, higher dimensionless frequencies preserve more geometric quantum discord between the accessible modes. This geometric quantum discord recovers its initial value in the Nariai limit, while geometric quantum discord in all three combinations containing inaccessible modes vanishes. At higher dimensionless frequencies, the unequal horizon temperatures suppress geometric quantum discord involving the inaccessible mode beyond the cosmological horizon. By contrast, competing thermal effects on the inaccessible mode inside the black hole and the accessible mode near the cosmological horizon produce crossings between their frequency curves. For the selected states with equal initial geometric quantum discord, the two families yield identical geometric quantum discord both between these latter modes and between the two inaccessible partners. Differences persist over part of the range between the accessible mode near the black hole horizon and the inaccessible mode beyond the cosmological horizon. Changes in the optimal local measurement account for the kinks in the curves. These results show how horizon asymmetry and the initial state jointly affect geometric quantum discord in different modes.

Sec.~II describes the SdS geometry and Dirac mode transformations at the two horizons. Sec.~III introduces geometric quantum discord and the initial states. Sec.~IV presents the reduced states and analyzes their correlations, and Sec.~V summarizes the results.

\section{II. Quantum thermal effects in Schwarzschild de Sitter spacetime}

We use natural units, $\hbar=c=G=k_B=1$. The Schwarzschild de Sitter (SdS) spacetime describes a static, spherically symmetric black hole of mass $M$ with a positive cosmological constant $\Lambda$. Its line element is given by \cite{Liu2023,Major2022,BhattacharyaJoshi2022,Bhattacharya2018}
\begin{equation}\label{eq:metric}
ds^2=-f(r)dt^2+\frac{dr^2}{f(r)}+r^2d\Omega^2,
\end{equation}
where $d\Omega^2=d\theta^2+\sin^2\theta\,d\phi^2$ denotes the metric on the unit two sphere. The metric function can be factorized as
\begin{equation}\label{eq:metricfunction}
\begin{aligned}
f(r)&=1-\frac{2M}{r}-\frac{\Lambda r^2}{3}\\
&=\frac{\Lambda}{3r}(r-r_H)(r_C-r)(r-r_U), 
\end{aligned}
\end{equation}
where $r_H$ and $r_C$ denote the radii of the black hole event horizon (BEH) and the cosmological event horizon (CEH), respectively, while $r_U=-(r_H+r_C)$ denotes the negative root and therefore lies outside the physical radial domain. The dimensionless combination $3M\sqrt{\Lambda}$ parametrizes the approach to coincident horizons. The regime with two distinct positive horizons satisfies
\begin{equation}\label{eq:nariai}
0<3M\sqrt{\Lambda}<1.
\end{equation}
Within this parameter range, the two positive roots of $f(r)=0$ are given by
\begin{equation}\label{eq:horizons}
\begin{aligned}
r_H&=\frac{2}{\sqrt{\Lambda}}
\cos\left[\frac{\arccos(-3M\sqrt{\Lambda})+4\pi}{3}\right],\\
r_C&=\frac{2}{\sqrt{\Lambda}}
\cos\left[\frac{\arccos(-3M\sqrt{\Lambda})}{3}\right].
\end{aligned}
\end{equation}
The inequality $f(r)>0$ identifies $r_H<r<r_C$ as the static region. At the Nariai limit, $3M\sqrt{\Lambda}=1$, the two horizons coincide at $r_H=r_C=1/\sqrt{\Lambda}$, corresponding to the largest black hole allowed for a fixed cosmological constant. The combination $3M\sqrt{\Lambda}$ will be used to characterize the dependence of quantum correlations on the spacetime geometry.

The surface gravities of the BEH and CEH are given by \cite{Liu2023,GibbonsHawking1977}

\begin{equation}\label{eq:surface}
\begin{aligned}
\kappa_H&=\frac{f'(r_H)}{2}
=\frac{\Lambda(r_C-r_H)(2r_H+r_C)}{6r_H},\\
\kappa_C&=-\frac{f'(r_C)}{2}
=\frac{\Lambda(r_C-r_H)(r_H+2r_C)}{6r_C}.
\end{aligned}
\end{equation}
The signed surface gravity is positive at the BEH and negative at the CEH, where $f'(r_C)/2=-\kappa_C$; here $\kappa_H$ and $\kappa_C$ denote positive magnitudes. As the Nariai limit is approached, both magnitudes and the corresponding horizon temperatures tend to zero with the normalization used here \cite{Liu2023}.

To extend the coordinates across the BEH and CEH, we introduce the tortoise coordinate \cite{BhattacharyaJoshi2022,Bhattacharya2018}
\begin{equation}\label{eq:tortoise}
\begin{aligned}
r_\star&=\int\frac{dr}{f(r)}\\
&=\frac{\ln|r/r_H-1|}{2\kappa_H}
-\frac{\ln|1-r/r_C|}{2\kappa_C}
+\frac{\ln|r/r_U-1|}{2\kappa_U},
\end{aligned}
\end{equation}
up to an additive constant, where $\kappa_U=f'(r_U)/2>0$ is associated with the negative root. Exponential null coordinates remove the logarithmic divergences in Eq.~(\ref{eq:tortoise}) locally. With $u=t-r_\star$ and $v=t+r_\star$, the transformations are given by \cite{BhattacharyaJoshi2022}
\begin{equation}\label{eq:kruskal}
\begin{aligned}
\bar u_H&=-\kappa_H^{-1}e^{-\kappa_H u},&
\bar v_H&=\kappa_H^{-1}e^{\kappa_H v},\\
\bar u_C&=\kappa_C^{-1}e^{\kappa_C u},&
\bar v_C&=-\kappa_C^{-1}e^{-\kappa_C v}.
\end{aligned}
\end{equation}
Substitution into Eq.~(\ref{eq:metric}) yields the following line elements:
\begin{equation}\label{eq:regularmetric}
\begin{aligned}
ds^2&=-f(r)e^{-2\kappa_H r_\star}\,d\bar u_H\,d\bar v_H+r^2d\Omega^2,\\
ds^2&=-f(r)e^{2\kappa_C r_\star}\,d\bar u_C\,d\bar v_C+r^2d\Omega^2.
\end{aligned}
\end{equation}
The two line elements are regular at the BEH and CEH, respectively. In the limit $\Lambda\to0$, one recovers $r_H\to2M$, $r_C\to\infty$ and the Schwarzschild geometry.

The two horizons generally have different temperatures and are not in thermal equilibrium. A thermally opaque membrane in the static region separates their neighborhoods into thermally isolated regions, so that each horizon can be treated independently \cite{Liu2023,BhattacharyaJoshi2022,Bhattacharya2018,Gomberoff2003,Sekiwa2006,Wu2024}. We consider modes with negligible transmission between these regions \cite{Liu2023}. Alice and Bob have access to modes $A$ and $B$ near the BEH and CEH, with temperatures $T_H=\kappa_H/(2\pi)$ and $T_C=\kappa_C/(2\pi)$, respectively. Mode $L$, inside the black hole event horizon, is the inaccessible partner of $A$; mode $R$, beyond the cosmological event horizon, is the inaccessible partner of $B$. Observers in the static region cannot access either partner.

To determine the vacuum and excited states in the two horizon neighborhoods, we consider a free massless Dirac field satisfying \cite{Alsing2006}

\begin{equation}\label{eq:dirac}
\gamma^a e_a{}^\mu(\partial_\mu+\Gamma_\mu)\Psi=0,
\end{equation}
where $\gamma^a$, $e_a{}^\mu$ and $\Gamma_\mu$ denote the Dirac matrices, the inverse tetrad and the spin connection, respectively.

We first consider the side of the membrane facing the BEH. Quantization in the local and Kruskal modes yields Bogoliubov transformations between the corresponding creation and annihilation operators, as in the treatment of a single black hole horizon \cite{Fuentes2005,Wang2010,Wang2014,Martin2010}. The same construction applies to the CEH. For a mode of frequency $\omega$, the fermionic mixing parameters at the BEH and CEH are given by \cite{Liu2023}

\begin{equation}\label{eq:dimensionless}
\tan r=e^{-\pi\omega/\kappa_H},\qquad
\tan s=e^{-\pi\omega/\kappa_C},
\end{equation}
where $r$ and $s$ denote mixing angles, with $r$ in Eq.~(\ref{eq:dimensionless}) distinguished from the radial coordinate. Their dependence on $3M\sqrt{\Lambda}$ follows from the horizon radii and surface gravities in Eqs.~(\ref{eq:horizons}) and (\ref{eq:surface}). We employ the single mode approximation and suppress the fixed spin and angular labels \cite{Liu2023,Alsing2006}. After normalization, the BEH Kruskal vacuum and its particle excitation can be expressed in the occupation basis of modes $A$ and $L$ as

\begin{equation}\label{eq:blackstates}
\begin{aligned}
|0\rangle_{\kappa_H}&=\cos r|0_A0_L\rangle
+\sin r|1_A1_L\rangle,\\
|1\rangle_{\kappa_H}&=|1_A0_L\rangle.
\end{aligned}
\end{equation}

For the CEH, the corresponding Kruskal vacuum and particle excitation are expanded in modes $B$ and $R$. With the same ordering convention, these states are given by

\begin{equation}\label{eq:cosmostates}
\begin{aligned}
|0\rangle_{\kappa_C}&=\cos s|0_B0_R\rangle
+\sin s|1_B1_R\rangle,\\
|1\rangle_{\kappa_C}&=|1_B0_R\rangle.
\end{aligned}
\end{equation}

\section{III. Geometric quantum discord and initial states}
\subsection{A. Geometric quantum discord}

Quantum discord, introduced by Ollivier and Zurek and by Henderson and Vedral \cite{OllivierZurek2001,HendersonVedral2001}, captures quantum correlations that can persist even in separable states. We use the geometric measure proposed by Daki\'c \emph{et al.} \cite{Dakic2010}, which can also be evaluated explicitly in relativistic settings \cite{LiHuang2026,Xiong2026}. It is defined as
\begin{equation}\label{eq:geometricdef}
D_G(\rho)=\min_{\chi\in\Omega_0}\|\rho-\chi\|_2^2,
\end{equation}
where $\Omega_0$ denotes the set of states with vanishing geometric quantum discord with respect to measurements on the first subsystem, and $\|\rho-\chi\|_2^2=\operatorname{Tr}[(\rho-\chi)^2]$ denotes the squared Hilbert Schmidt distance.

The identity and Pauli matrices form an orthogonal operator basis, so any two qubit state can be written in the Bloch representation as
\begin{equation}\label{eq:bloch}
\begin{aligned}
\rho&=\frac14\Bigl(I\otimes I+\sum_{i=1}^3 x_i\sigma_i\otimes I\\
&\quad+\sum_{j=1}^3 y_j I\otimes\sigma_j
+\sum_{i,j=1}^3 t_{ij}\sigma_i\otimes\sigma_j\Bigr),
\end{aligned}
\end{equation}
where $I$ denotes the identity on a single qubit and $\sigma_i$ are the Pauli matrices. The coefficients are given by $x_i=\operatorname{Tr}[\rho(\sigma_i\otimes I)]$, $y_j=\operatorname{Tr}[\rho(I\otimes\sigma_j)]$ and $t_{ij}=\operatorname{Tr}[\rho(\sigma_i\otimes\sigma_j)]$. In terms of the Bloch vector $\boldsymbol x=(x_1,x_2,x_3)^{\mathsf T}$ and the correlation matrix $T=(t_{ij})$, the geometric quantum discord can be evaluated as \cite{Dakic2010}
\begin{equation}\label{eq:discordclosed}
D_G(\rho)=\frac14\left(\|\boldsymbol x\|^2+\|T\|^2-\lambda_{\max}\right),
\end{equation}
where $\|\boldsymbol x\|$ denotes the Euclidean norm, while $\|T\|$ denotes the Hilbert Schmidt norm of the real correlation matrix, equivalently its Frobenius norm, with $\|T\|^2=\operatorname{Tr}(TT^{\mathsf T})$ \cite{Dakic2010}. The quantity $\lambda_{\max}$ denotes the largest eigenvalue of $\boldsymbol x\boldsymbol x^{\mathsf T}+TT^{\mathsf T}$, and the superscript $\mathsf T$ indicates transposition. 

\subsection{B. Bell type and isotropic initial states}

We consider two families of states encoded in the regular modes associated with the BEH and CEH. In this subsection, $|mn\rangle$ abbreviates $|m\rangle_{\kappa_H}\otimes|n\rangle_{\kappa_C}$ before applying the mode expansions of Sec.~II. The first family consists of pure Bell type states, whereas the second introduces isotropic mixing with the maximally mixed state.

The first family comprises pure Bell type states of the form \cite{LiShangWu2025}
\begin{equation}\label{eq:bellinput}
\rho_{\mathrm{Bell}}=|\psi_p\rangle\langle\psi_p|, 
\end{equation}
where $|\psi_p\rangle=p|01\rangle+\sqrt{1-p^2}|10\rangle$, with $0\leq p\leq1$. The populations of $|01\rangle$ and $|10\rangle$ are $p^2$ and $1-p^2$, respectively. The state remains pure as it changes from $|10\rangle$ at $p=0$, through the maximally entangled state $|\psi^+\rangle=(|01\rangle+|10\rangle)/\sqrt2$ at $p=1/\sqrt2$, to $|01\rangle$ at $p=1$. Its Bloch representation is given by
\begin{equation}\label{eq:bellbloch}
\begin{aligned}
\rho_{\mathrm{Bell}}&=\frac14\Bigl[I\otimes I+
(2p^2-1)(\sigma_3\otimes I-I\otimes\sigma_3)\\
&\quad+2p\sqrt{1-p^2}
(\sigma_1\otimes\sigma_1+\sigma_2\otimes\sigma_2)
-\sigma_3\otimes\sigma_3\Bigr].
\end{aligned}
\end{equation}
Consequently, $\boldsymbol x=(0,0,2p^2-1)^{\mathsf T}$, $\boldsymbol y=-\boldsymbol x$ and $T=\operatorname{diag}(2p\sqrt{1-p^2},2p\sqrt{1-p^2},-1)$. Except at the maximally entangled point, the local Bloch vectors do not vanish.

For comparison, we consider the two qubit isotropic family \cite{WangBattery2026,Horodecki1999}
\begin{equation}\label{eq:isoinput}
\rho_{\mathrm{iso}}=\frac{1-q}{4}I\otimes I
+q|\psi^+\rangle\langle\psi^+|.
\end{equation}
The initial state parameter $q$ ranges from zero to one and controls the weight of the entangled component: $q=0$ gives the maximally mixed state, while $q=1$ recovers $|\psi^+\rangle$. As $q$ increases, the purity $(1+3q^2)/4$ increases from $1/4$ to one, and the state becomes entangled for $q>1/3$. The isotropic state given in Eq.~(\ref{eq:isoinput}) has the Bloch representation
\begin{equation}\label{eq:isobloch}
\rho_{\mathrm{iso}}=\frac14\bigl[I\otimes I
+q(\sigma_1\otimes\sigma_1+\sigma_2\otimes\sigma_2
-\sigma_3\otimes\sigma_3)\bigr],
\end{equation}
with $\boldsymbol x=\boldsymbol y=0$ and $T=\operatorname{diag}(q,q,-q)$. Unlike the Bell type family, varying $q$ changes the global purity while leaving both local states maximally mixed.

Substituting the Bloch coefficients of the Bell type and isotropic states into Eq.~(\ref{eq:discordclosed}), we obtain their initial geometric quantum discords, respectively, as
\begin{equation}\label{eq:initialdiscord}
D_G(\rho_{\mathrm{Bell}})=2p^2(1-p^2),
\qquad D_G(\rho_{\mathrm{iso}})=\frac{q^2}{2}.
\end{equation}
These expressions show how the initial state parameters set the initial correlations. The geometric quantum discord of the Bell type state reaches its maximum at $p=1/\sqrt2$. This pure state is entangled and has nonzero geometric quantum discord for $0<p<1$; both vanish at the product state endpoints $p=0$ and $p=1$. In contrast, the geometric quantum discord of the isotropic state increases monotonically with $q$ and is nonzero for $q>0$, whereas entanglement requires $q>1/3$. Thus, the interval $0<q\leq1/3$ contains separable states with quantum correlations detected by geometric quantum discord. These states allow us to examine how different initial states respond to the thermal effects of the two horizons.

\section{IV. Reduced states across the two horizons}

\begin{figure}[ht]
\centering
\includegraphics[width=\columnwidth]{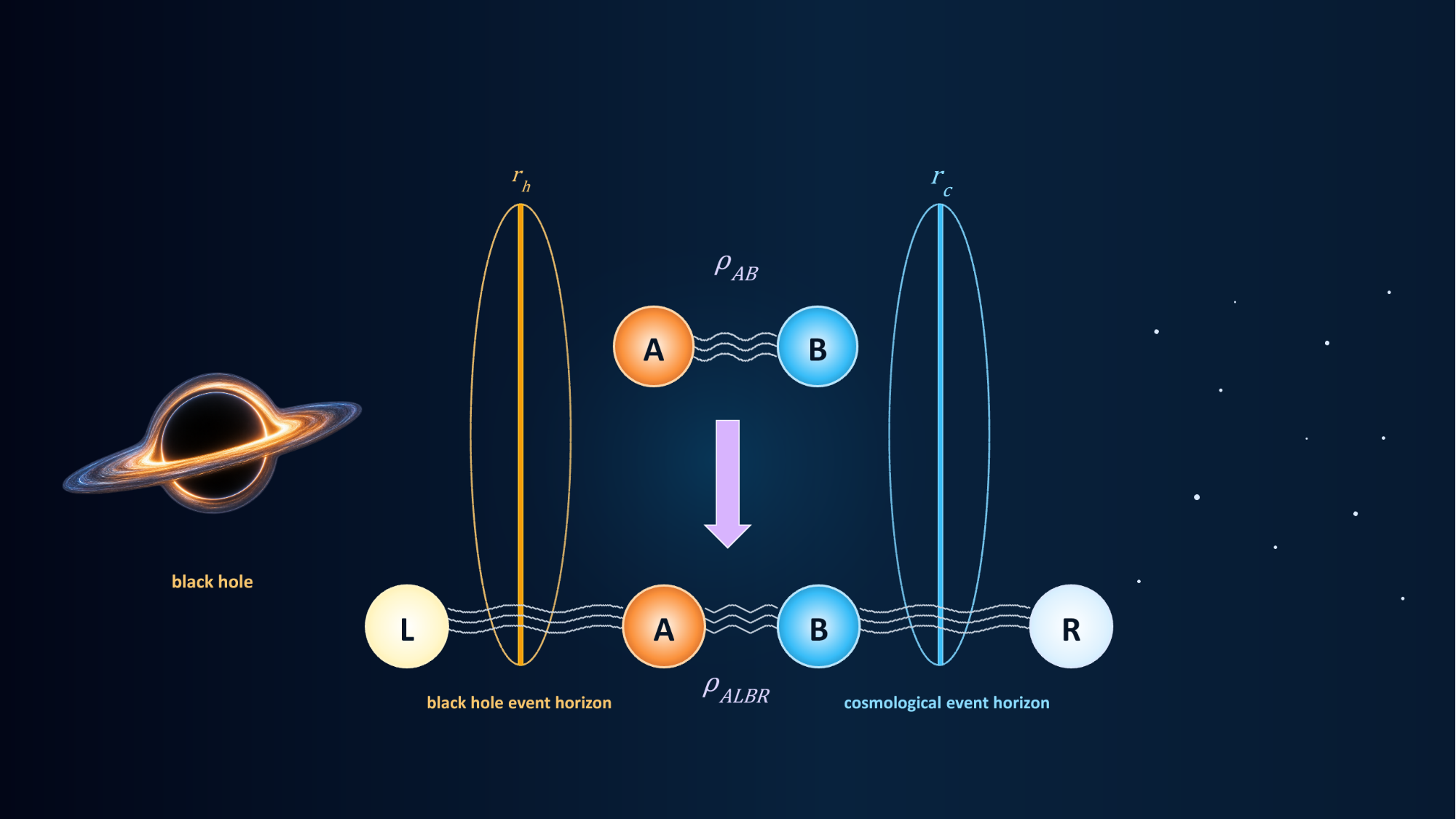}
\caption{Mode transformation from the initial state $\rho_{AB}$ to $\rho_{ALBR}$ in SdS spacetime. The BEH and CEH associate the initial modes with modes $AL$ and $BR$, respectively. Modes $A$ and $B$ are physically accessible, while $L$ and $R$ are inaccessible. Wavy lines schematically represent quantum correlations.}
\label{fig:mode-schematic}
\end{figure}

In this section, we examine how the two horizon effects change the correlations of the initial states introduced above. Suppose Alice and Bob initially share a state $\rho$, chosen as either $\rho_{\mathrm{Bell}}$ in Eq.~(\ref{eq:bellinput}) or $\rho_{\mathrm{iso}}$ in Eq.~(\ref{eq:isoinput}). The mode expansions in Eqs.~(\ref{eq:blackstates}) and (\ref{eq:cosmostates}) associate Alice's mode with $AL$ at the BEH and Bob's mode with $BR$ at the CEH, giving the four mode state $\rho_{ALBR}$. Fig.~\ref{fig:mode-schematic} illustrates this mode structure. Modes $A$ and $B$ remain accessible in the static region, while $L$ lies inside the BEH and $R$ lies beyond the CEH.

In the input basis $\{|0\rangle,|1\rangle\}$, the mode transformations in Eqs.~(\ref{eq:blackstates}) and (\ref{eq:cosmostates}) have the matrix representations
\begin{equation}\label{eq:modeisometries}
V_H=\begin{pmatrix}
\cos r&0\\0&0\\0&1\\\sin r&0
\end{pmatrix},\qquad
V_C=\begin{pmatrix}
\cos s&0\\0&0\\0&1\\\sin s&0
\end{pmatrix},
\end{equation}
where the rows follow the basis order $\{|00\rangle,|01\rangle,|10\rangle,|11\rangle\}$, with mode order $AL$ for $V_H$ and $BR$ for $V_C$. The matrices $V_H$ and $V_C$ preserve inner products, since $V_H^\dagger V_H=V_C^\dagger V_C=I$. Applying them to the initial state gives
\begin{equation}\label{eq:fourmodestate}
\rho_{ALBR}=(V_H\otimes V_C)\rho(V_H\otimes V_C)^\dagger,
\end{equation}
with output order $A,L,B,R$. Tracing over the unobserved modes then gives each two mode state. We first consider the accessible modes $AB$, followed by the inaccessible modes $AR$, $LB$ and $LR$.

\subsection{A. Physically accessible states}

Observers in the static region have access to $A$ and $B$, while the partner modes $L$ and $R$ are inaccessible. Tracing over these partners gives their shared state,
\begin{equation}\label{eq:accessibletrace}
\rho_{AB}=\operatorname{Tr}_{LR}(\rho_{ALBR}).
\end{equation}
For the Bell type initial state, evaluating this partial trace gives the Bloch representation
\begin{equation}\label{eq:bellAB}
\medmuskip=2mu \thinmuskip=2mu \thickmuskip=3mu
\begin{aligned}
\rho_{AB}^{\mathrm{Bell}}&=\frac14\Bigl[I\otimes I+(2p^2\cos^2r-1)\sigma_3\otimes I\\
&\quad+\bigl(2(1-p^2)\cos^2s-1\bigr)I\otimes\sigma_3\\
&\quad+2p\sqrt{1-p^2}\cos r\cos s
(\sigma_1\otimes\sigma_1+\sigma_2\otimes\sigma_2)\\
&\quad+\bigl(1-2p^2\cos^2r-2(1-p^2)\cos^2s\bigr)
\sigma_3\otimes\sigma_3\Bigr].
\end{aligned}
\end{equation}
For the isotropic initial state, the same partial trace gives
\begin{equation}\label{eq:isoAB}
\begin{aligned}
\rho_{AB}^{\mathrm{iso}}&=\frac14\Bigl[I\otimes I-\sin^2r\,\sigma_3\otimes I
-\sin^2s\,I\otimes\sigma_3\\
&\quad+q\cos r\cos s(\sigma_1\otimes\sigma_1+\sigma_2\otimes\sigma_2)\\
&\quad+(\sin^2r\sin^2s-q\cos^2r\cos^2s)\sigma_3\otimes\sigma_3\Bigr].
\end{aligned}
\end{equation}
Comparing Eqs.~(\ref{eq:bellAB}) and (\ref{eq:isoAB}) with the Bloch representation in Eq.~(\ref{eq:bloch}) gives the local Bloch vector $\boldsymbol x$ and correlation matrix $T$ for each reduced state.

Substituting these quantities into Eq.~(\ref{eq:discordclosed}), we obtain the geometric quantum discord of the Bell type reduced state as
\begin{equation}\label{eq:discordBellAB}
\begin{aligned}
D_G\bigl(\rho_{AB}^{\mathrm{Bell}}\bigr)
&=\frac14\min\Bigl\{8p^2(1-p^2)\cos^2r\cos^2s,\\
&\quad 4p^2(1-p^2)\cos^2r\cos^2s+(2p^2\cos^2r-1)^2\\
&\quad +\bigl(1-2p^2\cos^2r-2(1-p^2)\cos^2s\bigr)^2\Bigr\}.
\end{aligned}
\end{equation}
Similarly, the geometric quantum discord of the isotropic reduced state is
\begin{equation}\label{eq:discordIsoAB}
\begin{aligned}
D_G\bigl(\rho_{AB}^{\mathrm{iso}}\bigr)
&=\frac14\min\Bigl\{2q^2\cos^2r\cos^2s,\\
&\quad q^2\cos^2r\cos^2s+\sin^4r\\
&\quad +\bigl(\sin^2r\sin^2s-q\cos^2r\cos^2s\bigr)^2\Bigr\}.
\end{aligned}
\end{equation}
\begin{figure}[htbp]
\centering
\includegraphics[width=\columnwidth]{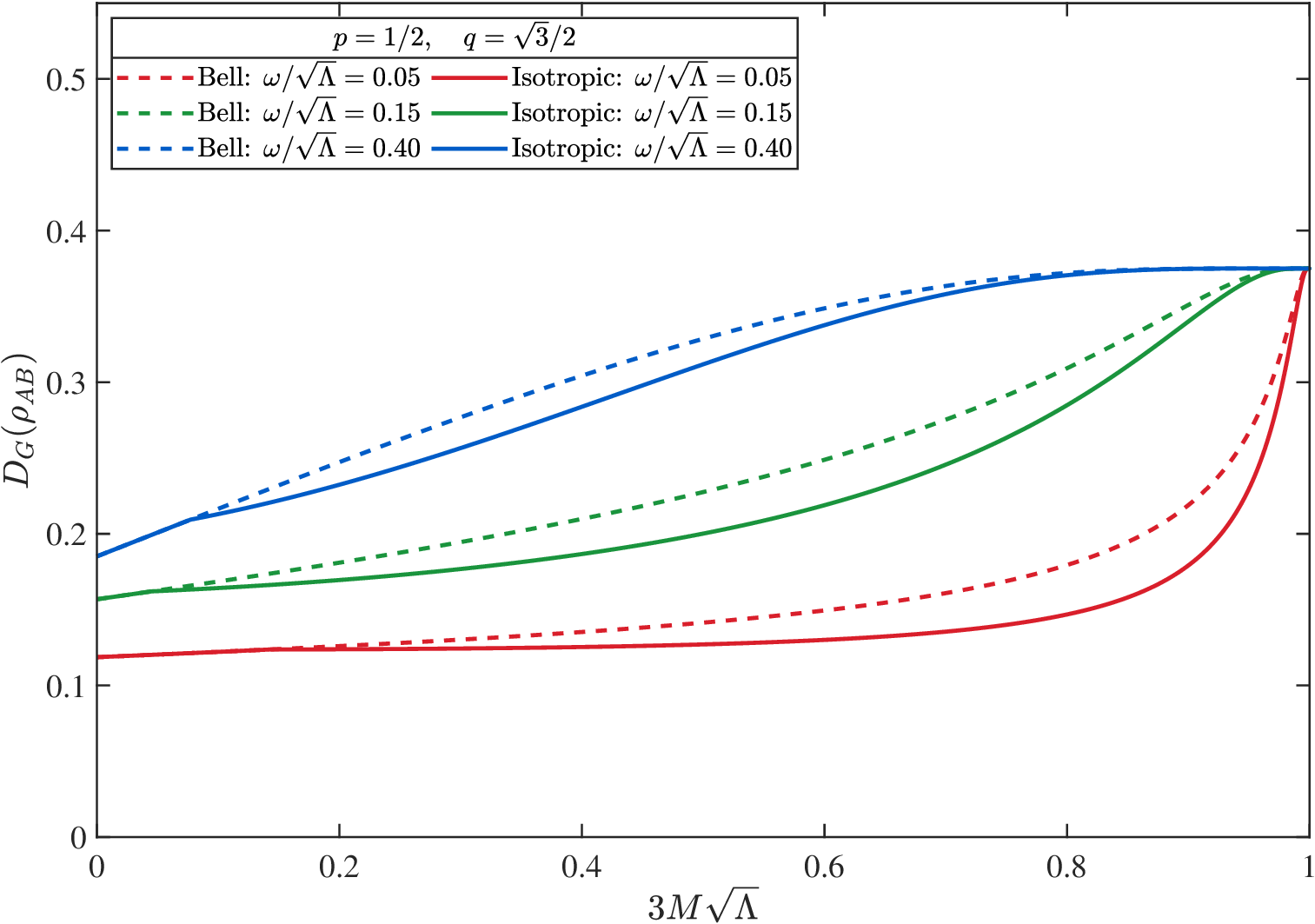}
\caption{Accessible geometric quantum discord for the Bell type initial state with $p=1/2$ and the isotropic initial state with $q=\sqrt3/2$. Both initial states have initial geometric quantum discord $0.375$. Red, green and blue lines correspond to dimensionless frequencies $\omega/\sqrt{\Lambda}=0.05$, $0.15$ and $0.40$, respectively. Dashed lines represent the Bell type initial state and solid lines the isotropic initial state. The horizontal endpoints denote limits of the family with two distinct horizons.}
\label{fig:AB-frequency}
\end{figure}

Fig.~\ref{fig:AB-frequency} shows the accessible geometric quantum discord for the Bell type state with $p=1/2$ and the isotropic state with $q=\sqrt3/2$. According to Eq.~(\ref{eq:initialdiscord}), both states have initial geometric quantum discord $0.375$. For all three dimensionless frequencies $\omega/\sqrt{\Lambda}=0.05$, $0.15$ and $0.40$, the accessible geometric quantum discord increases with $3M\sqrt{\Lambda}$. At fixed $\Lambda$, this corresponds to increasing the black hole mass $M$, which lowers both horizon temperatures and weakens the thermal effects on the accessible modes $AB$. As $3M\sqrt{\Lambda}\to1^{-}$, both states recover their initial geometric quantum discord.

At a given $3M\sqrt{\Lambda}$, a larger dimensionless frequency $\omega/\sqrt{\Lambda}$ gives a larger accessible geometric quantum discord. The blue curves approach the initial value well before the Nariai limit, whereas the red curves rise slowly over a broad interval and then increase steeply within a narrow region near $3M\sqrt{\Lambda}=1$. Their increase in this region is more rapid than that of the green and blue curves. Modes with smaller $\omega/\sqrt{\Lambda}$ remain more susceptible to thermal excitation and require lower horizon temperatures for their accessible geometric quantum discord to recover. The recovery of the red curves is therefore concentrated closer to the Nariai limit, where the green and blue curves have already nearly saturated. The red curves also flatten as they approach $0.375$.

For each dimensionless frequency, the solid and dashed curves initially coincide at small $3M\sqrt{\Lambda}$ and then separate, with the Bell type state retaining more accessible geometric quantum discord than the isotropic state. Their difference subsequently decreases, and both curves approach the same Nariai limit. This shows that different initial states can retain different amounts of accessible geometric quantum discord under the same horizon effects, even when their initial geometric quantum discord is equal. As both horizon temperatures tend to zero, these thermal effects vanish and the two states recover the same initial geometric quantum discord, causing the curves to converge.

\begin{figure}[htbp]
\centering
\includegraphics[width=\columnwidth]{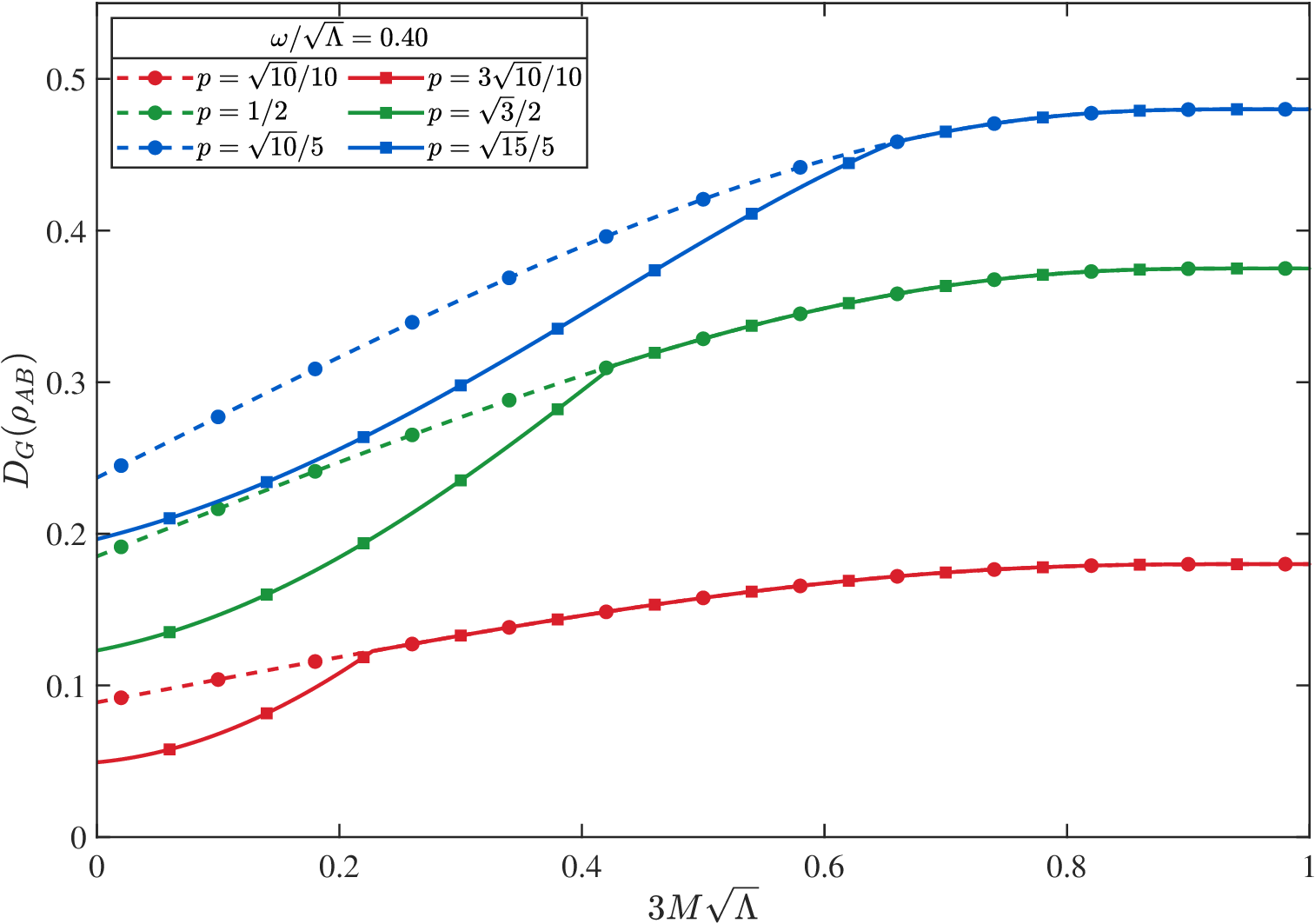}
\caption{Accessible geometric quantum discord for three pairs of Bell type initial states at dimensionless frequency $\omega/\sqrt{\Lambda}=0.40$. Red, green and blue lines correspond to $(p_1,p_2)=(\sqrt{10}/10,3\sqrt{10}/10)$, $(1/2,\sqrt3/2)$ and $(\sqrt{10}/5,\sqrt{15}/5)$, respectively. Each pair satisfies $p_1^2+p_2^2=1$, with $p_1<1/\sqrt2<p_2$. Dashed lines with circles denote $p_1$, and solid lines with squares denote $p_2$. The markers are interleaved to distinguish lines in their overlapping portions.}
\label{fig:AB-Bell-pairs}
\end{figure}

Fig.~\ref{fig:AB-Bell-pairs} compares three pairs of Bell type initial states at dimensionless frequency $\omega/\sqrt{\Lambda}=0.40$. The two values of $p$ in each pair lie on opposite sides of $1/\sqrt2$ and have squares that sum to unity. Eq.~(\ref{eq:initialdiscord}) therefore gives equal initial geometric quantum discord for the two states represented by each color, with values $0.18$, $0.375$ and $0.48$ for the red, green and blue pairs, respectively. All curves increase with the Nariai parameter $3M\sqrt{\Lambda}$ and approach their respective initial values in the Nariai limit. At fixed $\Lambda$, increasing the black hole mass $M$ lowers both horizon temperatures and weakens the thermal effects on the accessible modes.

For small Nariai parameters, the dashed curve with $p<1/\sqrt2$ lies above the solid curve with $p>1/\sqrt2$ in each pair. Thus, two Bell type initial states with equal initial geometric quantum discord can retain different amounts of accessible geometric quantum discord under the horizon effects. Replacing the smaller $p$ by the larger $p$ in a complementary pair interchanges the initial populations of $|01\rangle$ and $|10\rangle$. Since the black hole event horizon and the cosmological event horizon have different temperatures, these two initial states are affected differently. As the Nariai parameter increases, the solid curve meets the dashed curve of the same color at a kink, after which the two curves overlap.

The kinks occur at $3M\sqrt{\Lambda}\simeq0.227$, $0.427$ and $0.658$ for the red, green and blue pairs, respectively. As the paired values of $p$ approach $1/\sqrt2$, the onset of overlap shifts to larger Nariai parameters, extending the interval over which the two initial states yield different accessible geometric quantum discord. Beyond this point, choosing the smaller or larger $p$ gives the same accessible geometric quantum discord, although the corresponding reduced states need not be identical.

\begin{figure}[htbp]
\centering
\includegraphics[width=\columnwidth]{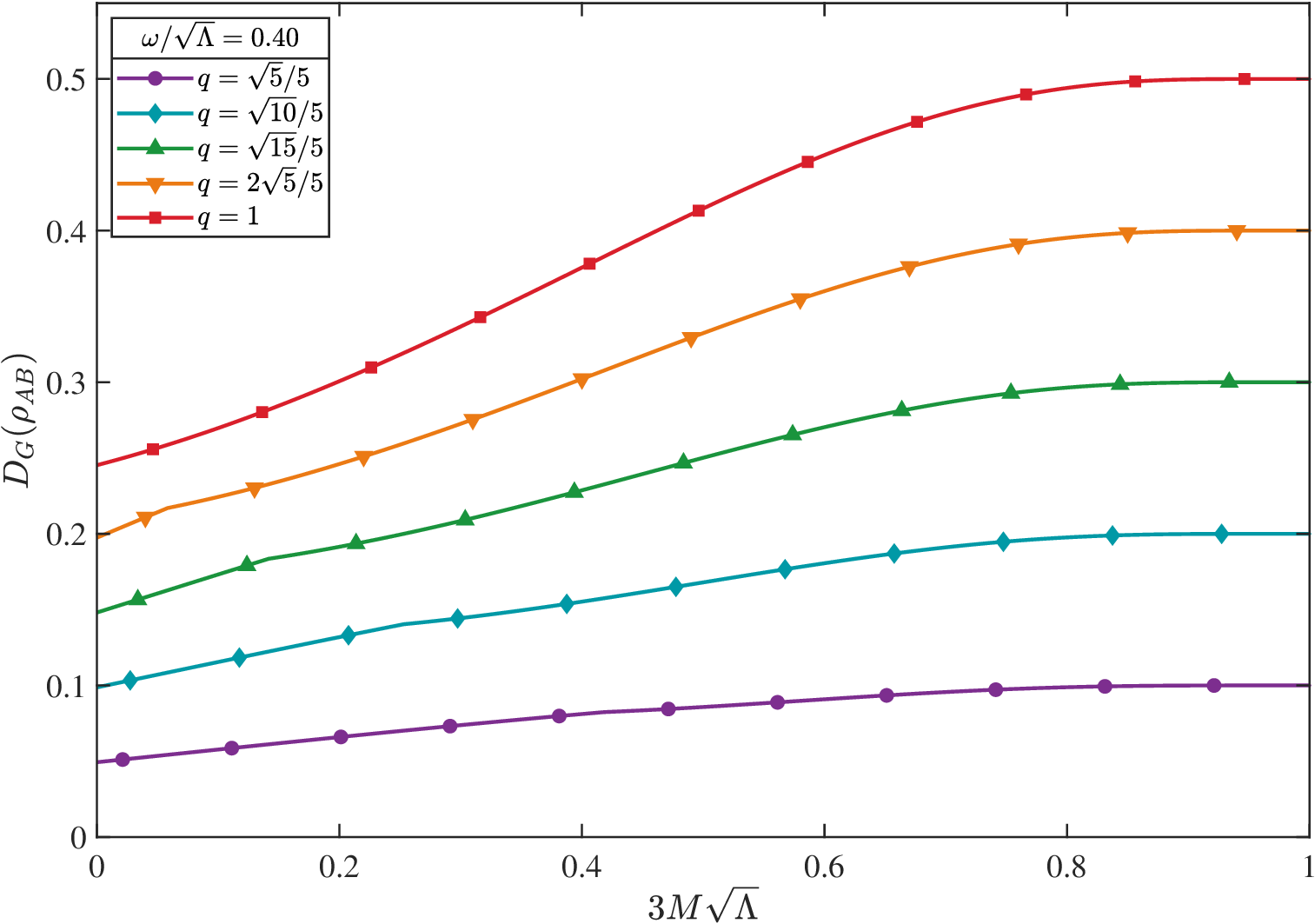}
\caption{Accessible geometric quantum discord for isotropic initial states at dimensionless frequency $\omega/\sqrt{\Lambda}=0.40$. Purple circles, cyan diamonds, green upward triangles, orange downward triangles and red squares correspond to $q=\sqrt5/5$, $\sqrt{10}/5$, $\sqrt{15}/5$, $2\sqrt5/5$ and $1$, respectively. These choices give initial geometric quantum discords $0.10$, $0.20$, $0.30$, $0.40$ and $0.50$.}
\label{fig:AB-isotropic}
\end{figure}

Fig.~\ref{fig:AB-isotropic} shows how the initial state parameter $q$ affects the accessible geometric quantum discord of isotropic states at dimensionless frequency $\omega/\sqrt{\Lambda}=0.40$. The purple, cyan, green, orange and red curves correspond to $q=\sqrt5/5$, $\sqrt{10}/5$, $\sqrt{15}/5$, $2\sqrt5/5$ and $1$, respectively. According to Eq.~(\ref{eq:initialdiscord}), these states have initial geometric quantum discord $0.10$, $0.20$, $0.30$, $0.40$ and $0.50$, respectively. All curves rise with the Nariai parameter $3M\sqrt{\Lambda}$ and approach their respective initial values in the Nariai limit. At fixed $\Lambda$, increasing the black hole mass $M$ lowers both horizon temperatures and weakens the thermal effects, allowing the accessible geometric quantum discord to recover.

At a given Nariai parameter, a larger $q$ yields greater accessible geometric quantum discord, and the curves retain this ordering throughout the plotted range. Increasing $q$ raises the weight of the Bell state and reduces that of the maximally mixed state, thereby increasing the initial geometric quantum discord. This ordering is preserved under the same horizon conditions. The uppermost curve corresponds to $q=1$, for which the initial state is a pure Bell state. The distinct limiting values reflect the different initial geometric quantum discord of these states.

The four curves with $q<1$ exhibit kinks that shift to smaller Nariai parameters as $q$ increases, whereas the red curve with $q=1$ varies smoothly throughout the plotted interval. These kinks correspond to changes in the optimal local measurement for geometric quantum discord. Thus, the initial state parameter $q$ affects both the accessible geometric quantum discord retained under the horizon effects and the location of the kink as the black hole mass varies.

\subsection{B. Physically inaccessible states}

The inaccessible modes $AR$ comprise the accessible mode near the black hole horizon and the inaccessible mode beyond the cosmological horizon. The inaccessible modes $LB$ comprise the inaccessible mode inside the black hole and the accessible mode near the cosmological horizon, while $LR$ contains the two inaccessible partners. The reduced states of these mode combinations follow by tracing over the complementary modes:
\begin{equation}\label{eq:inaccessibletraces}
\begin{aligned}
\rho_{AR}&=\operatorname{Tr}_{LB}(\rho_{ALBR}),\\
\rho_{LB}&=\operatorname{Tr}_{AR}(\rho_{ALBR}),\\
\rho_{LR}&=\operatorname{Tr}_{AB}(\rho_{ALBR}).
\end{aligned}
\end{equation}
We evaluate these traces using the four mode state in Eq.~(\ref{eq:fourmodestate}) for each initial state.

For the Bell type initial state, retaining $A$ and $R$ gives
\begin{equation}\label{eq:bellAR}
\medmuskip=2mu \thinmuskip=2mu \thickmuskip=3mu
\begin{aligned}
\rho_{AR}^{\mathrm{Bell}}&=\frac14\Bigl[I\otimes I+(2p^2\cos^2r-1)\sigma_3\otimes I\\
&\quad+\bigl(1-2(1-p^2)\sin^2s\bigr)I\otimes\sigma_3\\
&\quad+2p\sqrt{1-p^2}\cos r\sin s
(\sigma_1\otimes\sigma_1-\sigma_2\otimes\sigma_2)\\
&\quad+\bigl(2p^2\cos^2r+2(1-p^2)\sin^2s-1\bigr)
\sigma_3\otimes\sigma_3\Bigr].
\end{aligned}
\end{equation}
Retaining the complementary modes $L$ and $B$ instead yields
\begin{equation}\label{eq:bellLB}
\medmuskip=2mu \thinmuskip=2mu \thickmuskip=3mu
\begin{aligned}
\rho_{LB}^{\mathrm{Bell}}&=\frac14\Bigl[I\otimes I+(1-2p^2\sin^2r)\sigma_3\otimes I\\
&\quad+\bigl(2(1-p^2)\cos^2s-1\bigr)I\otimes\sigma_3\\
&\quad+2p\sqrt{1-p^2}\sin r\cos s
(\sigma_1\otimes\sigma_1-\sigma_2\otimes\sigma_2)\\
&\quad+\bigl(2p^2\sin^2r+2(1-p^2)\cos^2s-1\bigr)
\sigma_3\otimes\sigma_3\Bigr].
\end{aligned}
\end{equation}
When both accessible modes are discarded, the state shared by $L$ and $R$ takes the form
\begin{equation}\label{eq:bellLR}
\medmuskip=2mu \thinmuskip=2mu \thickmuskip=3mu
\begin{aligned}
\rho_{LR}^{\mathrm{Bell}}&=\frac14\Bigl[I\otimes I+(1-2p^2\sin^2r)\sigma_3\otimes I\\
&\quad+\bigl(1-2(1-p^2)\sin^2s\bigr)I\otimes\sigma_3\\
&\quad+2p\sqrt{1-p^2}\sin r\sin s
(\sigma_1\otimes\sigma_1+\sigma_2\otimes\sigma_2)\\
&\quad+\bigl(1-2p^2\sin^2r-2(1-p^2)\sin^2s\bigr)
\sigma_3\otimes\sigma_3\Bigr].
\end{aligned}
\end{equation}

For the isotropic initial state, the corresponding partial traces give
\begin{equation}\label{eq:isoAR}
\begin{aligned}
\rho_{AR}^{\mathrm{iso}}&=\frac14\Bigl[I\otimes I-\sin^2r\,\sigma_3\otimes I
+\cos^2s\,I\otimes\sigma_3\\
&\quad+q\cos r\sin s(\sigma_1\otimes\sigma_1-\sigma_2\otimes\sigma_2)\\
&\quad+(q\cos^2r\sin^2s-\sin^2r\cos^2s)\sigma_3\otimes\sigma_3\Bigr],
\end{aligned}
\end{equation}
\begin{equation}\label{eq:isoLB}
\begin{aligned}
\rho_{LB}^{\mathrm{iso}}&=\frac14\Bigl[I\otimes I+\cos^2r\,\sigma_3\otimes I
-\sin^2s\,I\otimes\sigma_3\\
&\quad+q\sin r\cos s(\sigma_1\otimes\sigma_1-\sigma_2\otimes\sigma_2)\\
&\quad+(q\sin^2r\cos^2s-\cos^2r\sin^2s)\sigma_3\otimes\sigma_3\Bigr],
\end{aligned}
\end{equation}
and
\begin{equation}\label{eq:isoLR}
\begin{aligned}
\rho_{LR}^{\mathrm{iso}}&=\frac14\Bigl[I\otimes I+\cos^2r\,\sigma_3\otimes I
+\cos^2s\,I\otimes\sigma_3\\
&\quad+q\sin r\sin s(\sigma_1\otimes\sigma_1+\sigma_2\otimes\sigma_2)\\
&\quad+(\cos^2r\cos^2s-q\sin^2r\sin^2s)\sigma_3\otimes\sigma_3\Bigr].
\end{aligned}
\end{equation}
For each ordered mode combination, substituting the local Bloch vector of its first mode and the diagonal correlation matrix into Eq.~(\ref{eq:discordclosed}) yields the corresponding geometric quantum discord.

\begin{figure}[tp]
\centering
\includegraphics[width=\columnwidth]{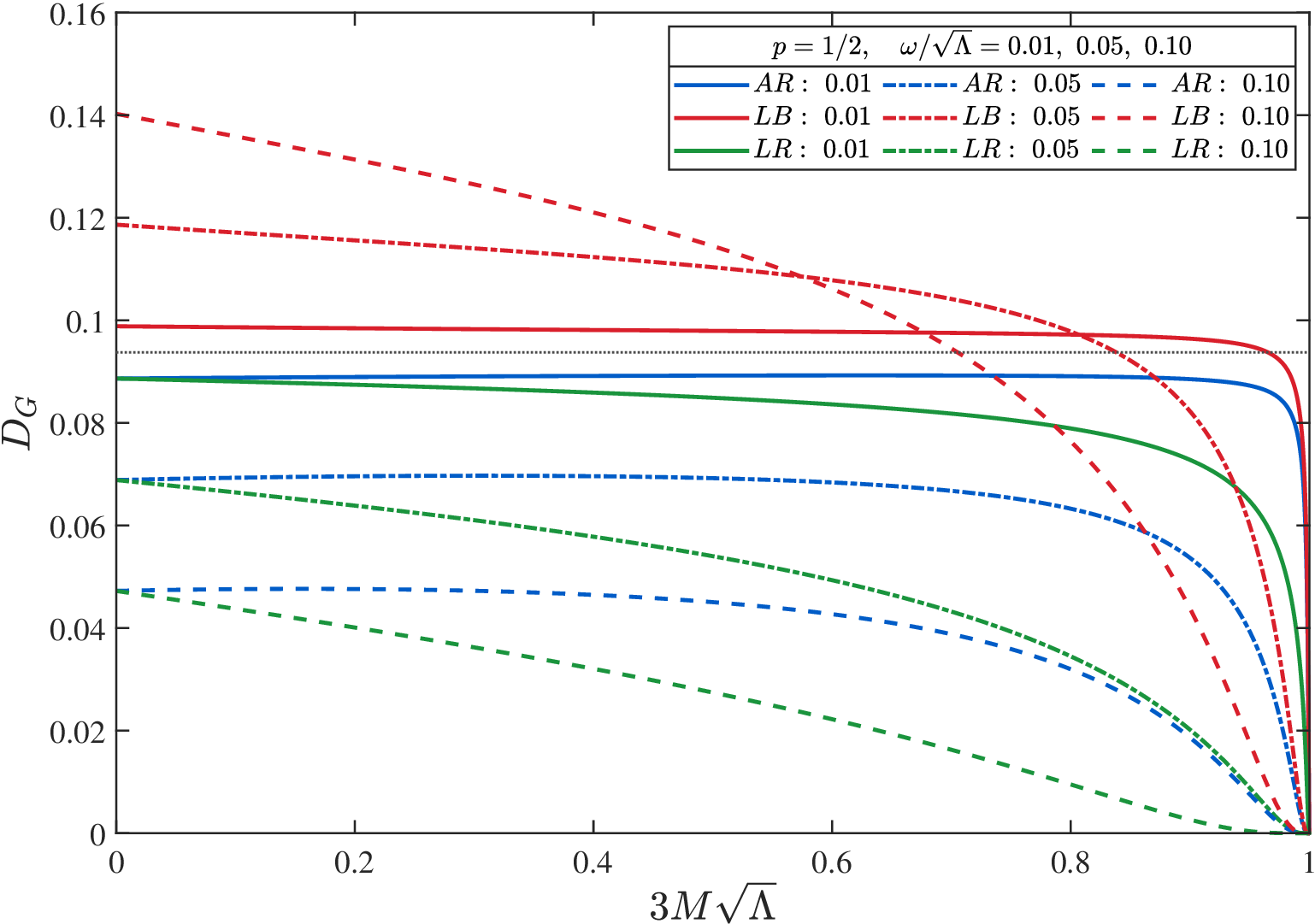}
\caption{Inaccessible geometric quantum discord for the Bell type initial state with $p=1/2$. Blue, red and green curves represent the inaccessible modes $AR$, $LB$ and $LR$, respectively. Solid, dash-dotted and dashed curves correspond to dimensionless frequencies $\omega/\sqrt{\Lambda}=0.01$, $0.05$ and $0.10$, respectively. The black dotted horizontal line marks the low frequency limit $0.09375$ for $p=1/2$ at fixed nondegenerate geometry. The horizontal endpoints represent limits of the geometry with two distinct horizons.}
\label{fig:inacc-modes}
\end{figure}

\begin{figure*}[tp]
\centering
\begin{minipage}[t]{0.325\textwidth}
\centering
\includegraphics[width=\linewidth]{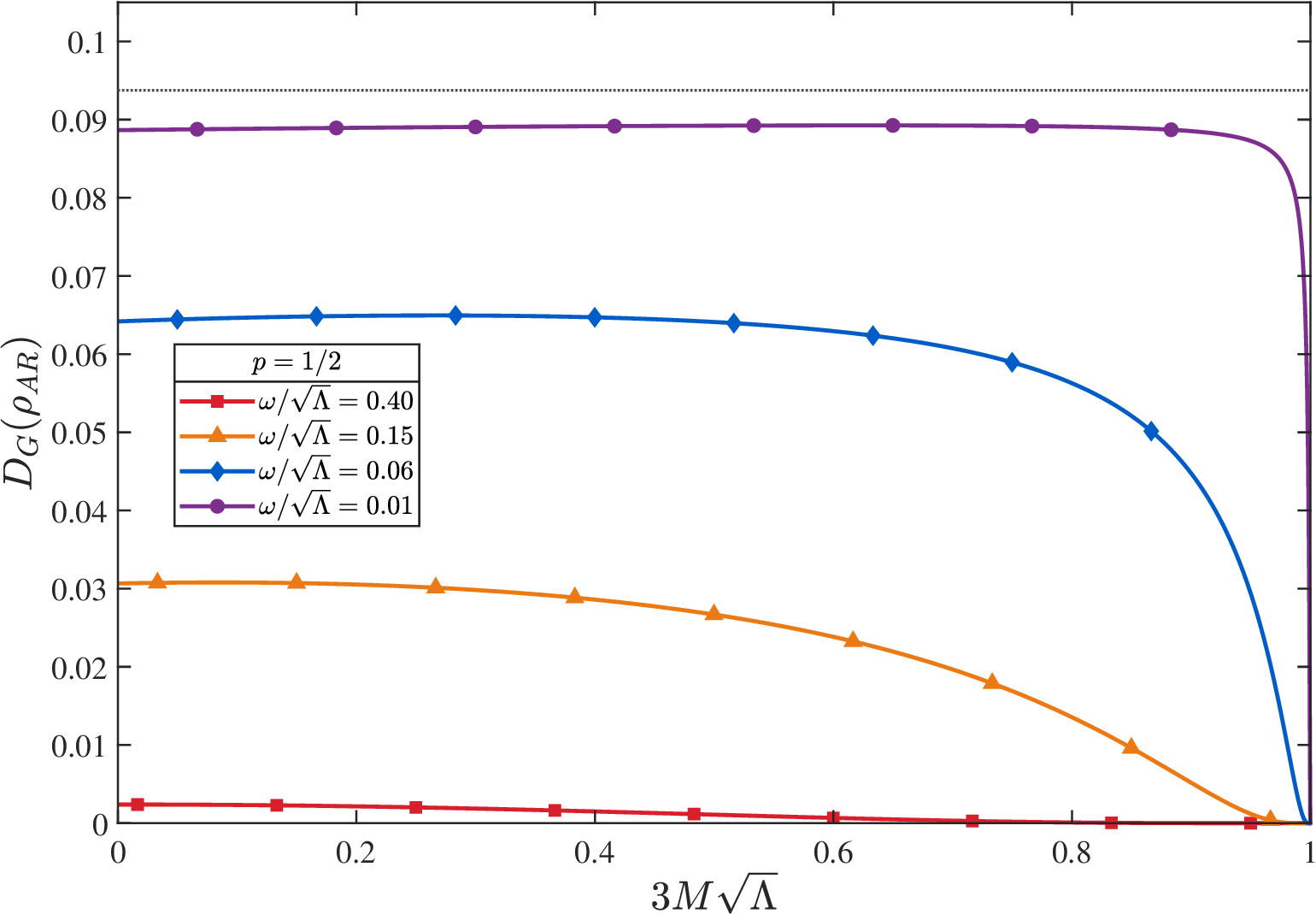}
\par{\small (a)}
\end{minipage}\hfill%
\begin{minipage}[t]{0.325\textwidth}
\centering
\includegraphics[width=\linewidth]{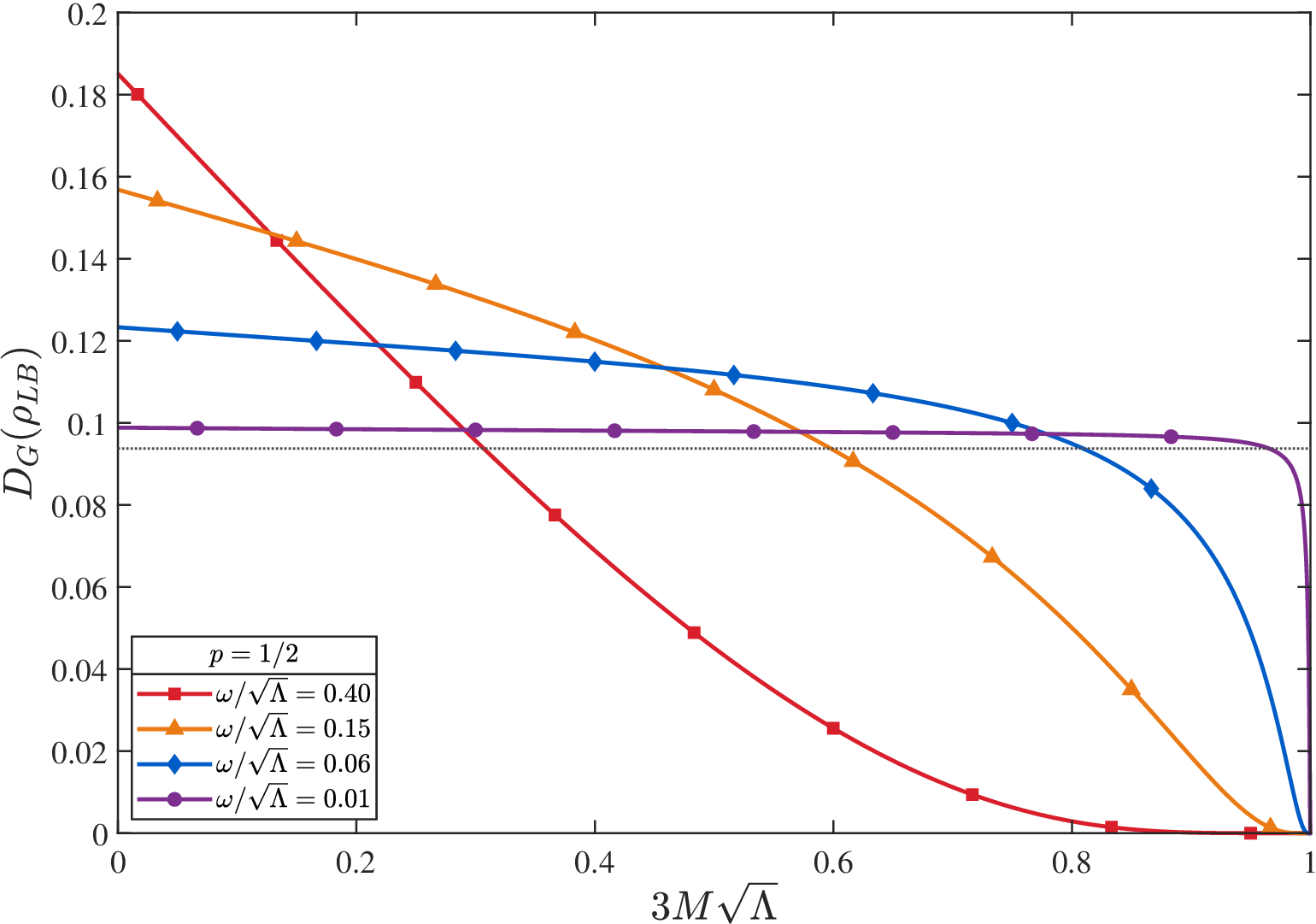}
\par{\small (b)}
\end{minipage}\hfill%
\begin{minipage}[t]{0.325\textwidth}
\centering
\includegraphics[width=\linewidth]{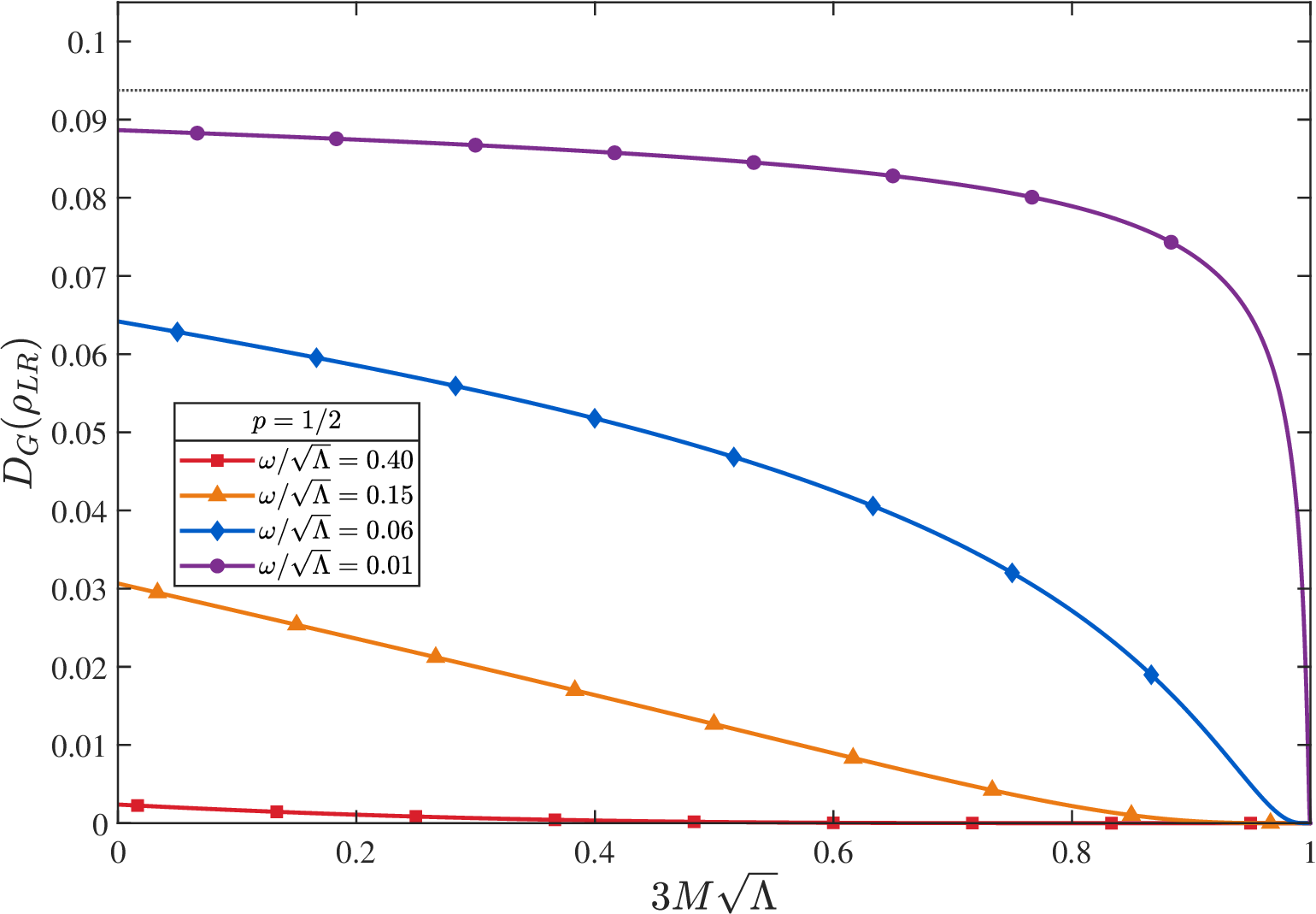}
\par{\small (c)}
\end{minipage}
\caption{Inaccessible geometric quantum discord for the Bell type initial state with $p=1/2$: (a) $AR$, (b) $LB$ and (c) $LR$. Red squares, orange triangles, blue diamonds and purple circles correspond to dimensionless frequencies $\omega/\sqrt{\Lambda}=0.40$, $0.15$, $0.06$ and $0.01$, respectively. The black dotted horizontal line marks $0.09375$. The vertical scales differ between panels.}
\label{fig:inacc-frequency}
\end{figure*}

\begin{figure*}[tp]
\centering
\begin{minipage}[t]{0.325\textwidth}
\centering
\includegraphics[width=\linewidth]{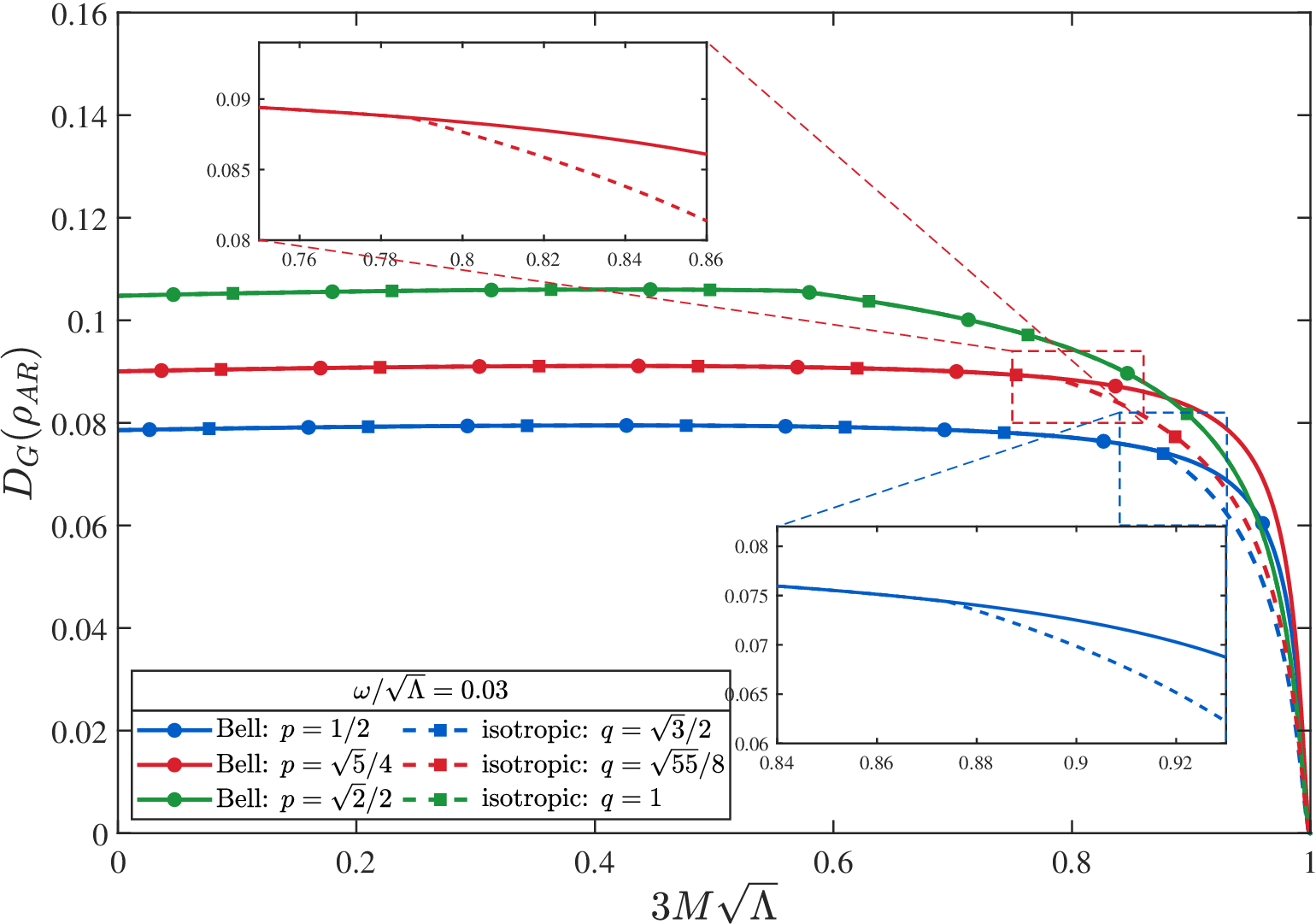}
\par{\small (a)}
\end{minipage}\hfill%
\begin{minipage}[t]{0.325\textwidth}
\centering
\includegraphics[width=\linewidth]{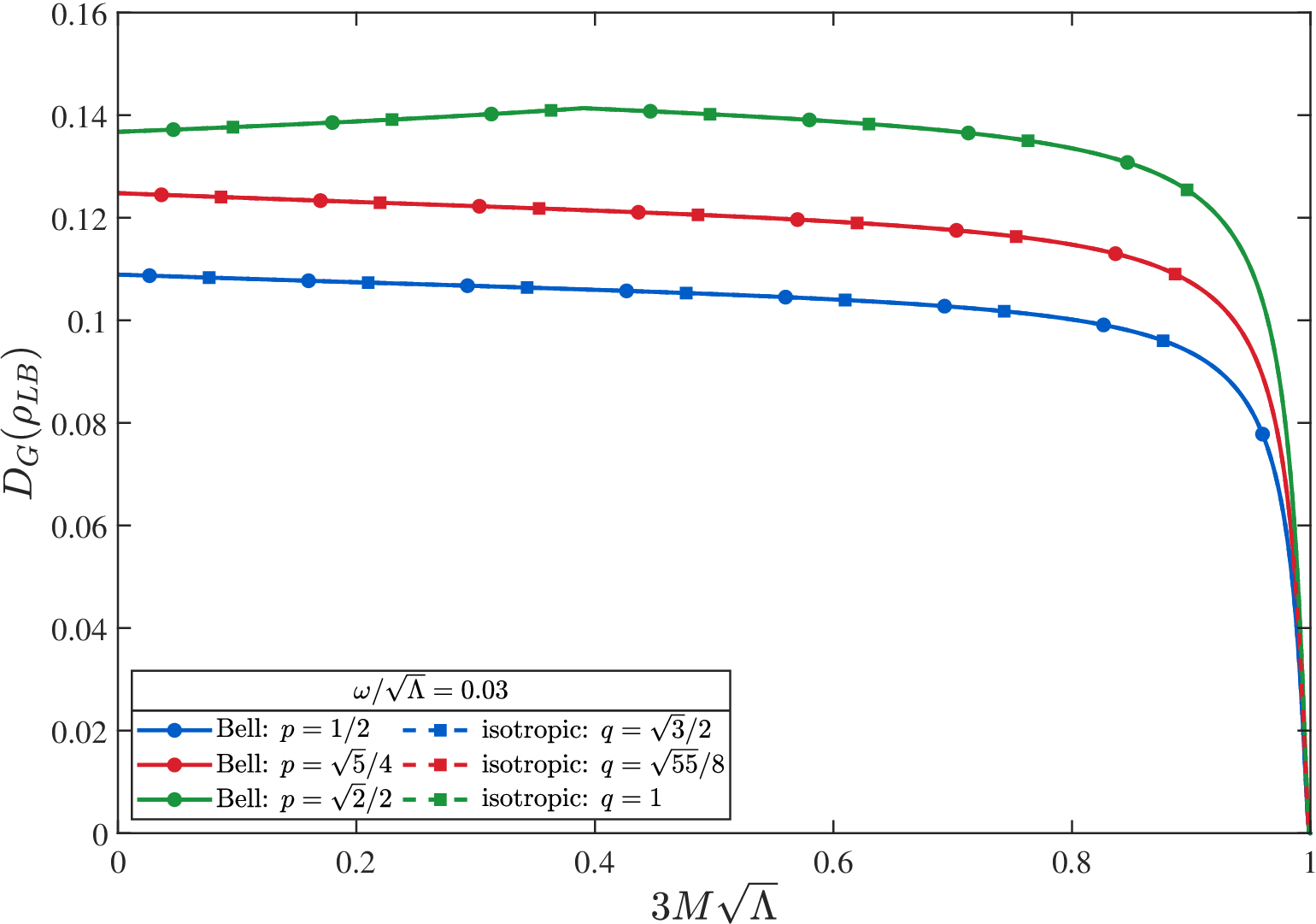}
\par{\small (b)}
\end{minipage}\hfill%
\begin{minipage}[t]{0.325\textwidth}
\centering
\includegraphics[width=\linewidth]{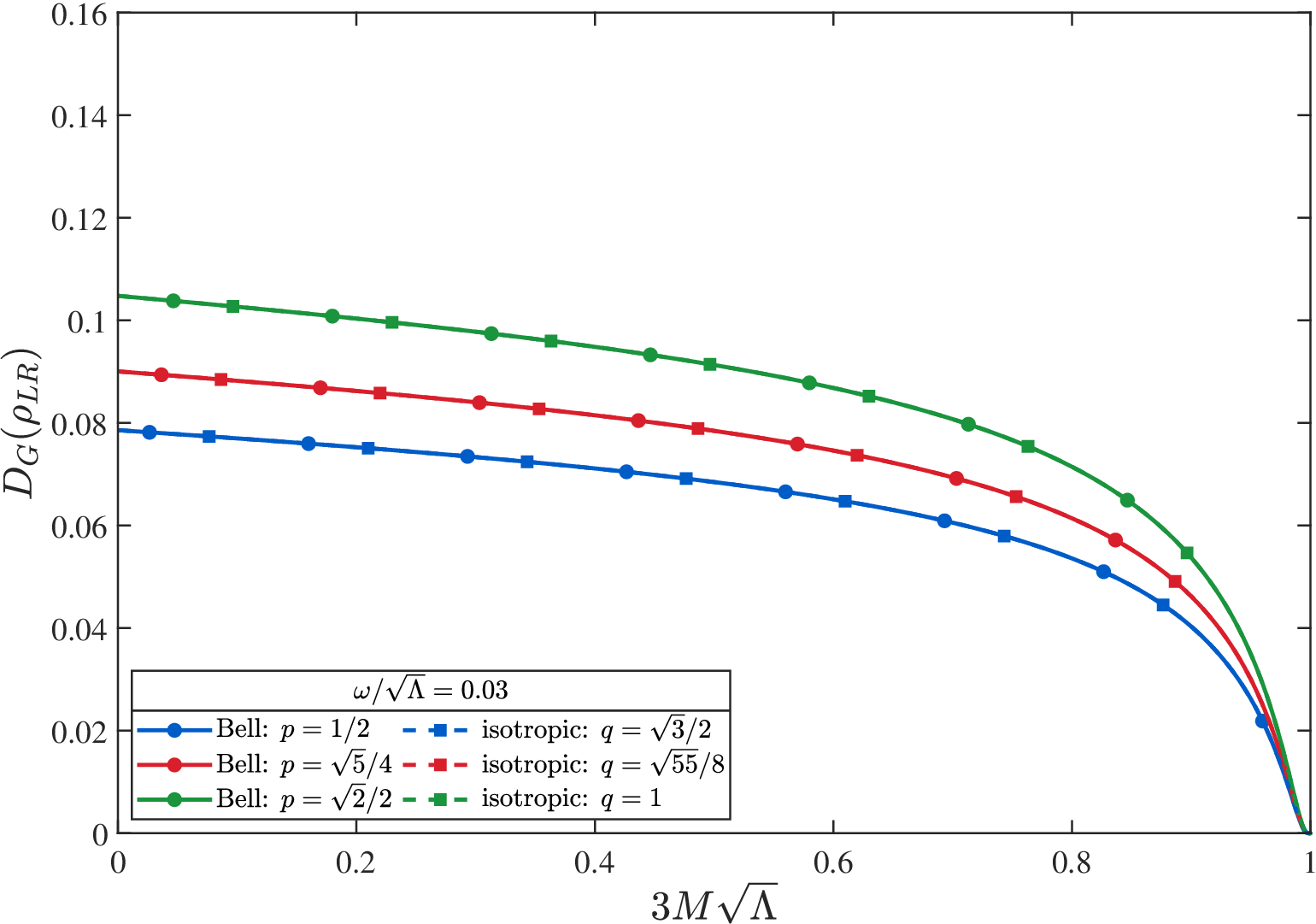}
\par{\small (c)}
\end{minipage}
\caption{Comparison of Bell type and isotropic initial states at dimensionless frequency $\omega/\sqrt{\Lambda}=0.03$: (a) $AR$, (b) $LB$ and (c) $LR$. Blue curves correspond to $p=1/2$ and $q=\sqrt3/2$, red curves to $p=\sqrt5/4$ and $q=\sqrt{55}/8$, and green curves to $p=\sqrt2/2$ and $q=1$. Each pair has equal initial geometric quantum discord, with values $0.375$, $0.4296875$ and $0.50$, respectively. Solid curves with filled circles denote Bell type states, and dashed curves with filled squares denote isotropic states. The upper red and lower blue insets in (a) enlarge the onset of separation in the corresponding boxed regions.}
\label{fig:inacc-states}
\end{figure*}

Fig.~\ref{fig:inacc-modes} through Fig.~\ref{fig:inacc-states} compare the inaccessible modes, their dependence on dimensionless frequency, and the two initial state families with matched initial geometric quantum discord.

Fig.~\ref{fig:inacc-modes} compares the inaccessible geometric quantum discord for the Bell type initial state with $p=1/2$. Blue, red and green curves represent the inaccessible modes $AR$, $LB$ and $LR$, respectively. For $p=1/2$ at fixed nondegenerate geometry, the geometric quantum discord of all three reduced states approaches $0.09375$ as the dimensionless frequency $\omega/\sqrt{\Lambda}\to0^{+}$, as obtained from Eq.~(\ref{eq:discordclosed}) and indicated by the black dotted horizontal line. Solid, dash-dotted and dashed curves correspond to dimensionless frequencies $0.01$, $0.05$ and $0.10$, respectively.

The inaccessible mode inside the black hole and the accessible mode near the cosmological horizon form $LB$. The inaccessible mode beyond the cosmological horizon combines with the accessible black hole side mode in $AR$, and with the inaccessible mode inside the black hole in $LR$. In the left half of the plotted interval, increasing the dimensionless frequency from $0.01$ through $0.05$ to $0.10$ raises the $LB$ geometric quantum discord and lowers those of $AR$ and $LR$. All three therefore move away from the black dotted reference line: $LB$ lies above it, while $AR$ and $LR$ lie below it. At each of these dimensionless frequencies, the $AR$ and $LR$ curves start at the same value in the left endpoint limit $3M\sqrt{\Lambda}\to0^{+}$ and separate as the Nariai parameter increases. The $LR$ geometric quantum discord subsequently falls below that of $AR$.

These different frequency responses reflect the unequal thermal effects of the two horizons. At small Nariai parameters, the high black hole event horizon temperature sustains the contribution of the partner mode $L$, while increasing dimensionless frequency reduces the thermal disturbance of the accessible mode $B$. The $LB$ geometric quantum discord consequently increases. The same frequency increase suppresses the contribution of the cosmological horizon partner $R$, lowering the geometric quantum discord of both $AR$ and $LR$. At fixed $\Lambda$, increasing the black hole mass $M$ lowers both horizon temperatures. The common low frequency value is obtained at fixed nondegenerate geometry. Along each plotted curve, the dimensionless frequency remains fixed and positive, and all three inaccessible geometric quantum discords approach zero in the Nariai limit.

Fig.~\ref{fig:inacc-frequency} shows the inaccessible geometric quantum discord of the Bell type initial state with $p=1/2$ for $AR$, $LB$ and $LR$ in panels (a), (b) and (c), respectively. Purple, blue, orange and red curves correspond to dimensionless frequencies $\omega/\sqrt{\Lambda}=0.01$, $0.06$, $0.15$ and $0.40$, respectively. The black dotted horizontal line indicates the low frequency limit $0.09375$ discussed above.

The accessible black hole side mode and the inaccessible mode beyond the cosmological horizon form $AR$, while the two inaccessible partners form $LR$. For both mode combinations, increasing the dimensionless frequency lowers the geometric quantum discord, and the ordering of the frequency curves is preserved. Their dependence on the Nariai parameter nevertheless differs. As the black hole mass $M$ increases at fixed $\Lambda$, the low frequency curves of $AR$ rise slightly before decreasing, whereas those of $LR$ decrease throughout. This difference reflects the modes retained in each combination. As both horizon temperatures fall, the reduced thermal disturbance of the accessible mode $A$ favors the preservation of correlations in $AR$, while the thermal contribution of the inaccessible partner $R$ decreases. The competition between these effects produces the shallow maximum in $AR$. In $LR$, which contains two inaccessible partners, the weakening thermal effects at both horizons jointly suppress the geometric quantum discord.

Fig.~\ref{fig:inacc-frequency}(b) concerns the inaccessible modes $LB$, formed by the inaccessible mode inside the black hole and the accessible mode near the cosmological horizon. The $LB$ curves exhibit a reversal of their frequency ordering. At small Nariai parameters, the geometric quantum discord increases as the dimensionless frequency rises from $0.01$ to $0.40$. As the black hole mass $M$ increases, the higher frequency curves decay more rapidly and cross the lower frequency curves. At larger Nariai parameters, their ordering is reversed, with lower dimensionless frequencies yielding larger geometric quantum discord. This behavior reflects competition between the reduced thermal disturbance of $B$ and the declining thermal contribution of the inaccessible mode $L$ at the BEH. At small black hole masses, the high BEH temperature sustains the contribution of $L$, so increasing the dimensionless frequency enhances the geometric quantum discord primarily by reducing the thermal disturbance of $B$. As the mass increases and the BEH cools, the contribution of $L$ is more strongly suppressed at higher frequencies, leading to the reversal.

For each fixed positive dimensionless frequency shown, all three inaccessible geometric quantum discords approach zero in the Nariai limit. At lower frequencies, appreciable geometric quantum discord persists over a wider range of the Nariai parameter, with the rapid decline confined closer to $3M\sqrt{\Lambda}=1$.

Fig.~\ref{fig:inacc-states} compares the inaccessible geometric quantum discord for Bell type and isotropic initial states at dimensionless frequency $\omega/\sqrt{\Lambda}=0.03$. Blue curves correspond to $p=1/2$ and $q=\sqrt3/2$, red curves to $p=\sqrt5/4$ and $q=\sqrt{55}/8$, and green curves to $p=\sqrt2/2$ and $q=1$. According to Eq.~(\ref{eq:initialdiscord}), the initial states represented by the same color have equal initial geometric quantum discord, with values $0.375$, $0.4296875$ and $0.50$, respectively. Solid and dashed curves denote Bell type and isotropic initial states, respectively.

Fig.~\ref{fig:inacc-states}(a) concerns $AR$, comprising the accessible mode near the black hole horizon and the inaccessible mode beyond the cosmological horizon. The green curves correspond to the same maximally entangled Bell state and therefore overlap throughout. The blue and red solid and dashed curves overlap at smaller Nariai parameters and separate as the black hole mass $M$ increases. After separation, the Bell type initial state yields greater geometric quantum discord than the isotropic initial state. The insets show the onset of separation, which occurs earlier for the red curves than for the blue curves. Both initial states ultimately yield vanishing geometric quantum discord in the Nariai limit. Equal initial geometric quantum discord can thus lead to different geometric quantum discord under the thermal effects of both horizons, with the onset of the difference depending on the initial state parameters.

In Fig.~\ref{fig:inacc-states}(b) and (c), the inaccessible mode inside the black hole is combined with the accessible cosmological side mode in $LB$ and with the inaccessible mode beyond the cosmological horizon in $LR$. For both $LB$ and $LR$, the solid and dashed curves of each color overlap throughout the interval for the three parameter choices considered. The influence of the initial state on inaccessible geometric quantum discord therefore depends on the modes retained. In both panels, the green, red and blue curves appear in descending order, consistent with the ordering of their initial geometric quantum discord. In $AR$, curves of different colors cross at larger Nariai parameters, showing that the initial geometric quantum discord and the initial state jointly affect the resulting geometric quantum discord. For all the initial states considered, the geometric quantum discord of the three inaccessible mode combinations vanishes in the Nariai limit.

\section{V. Conclusions}

We have investigated the geometric quantum discord of Dirac field modes in Schwarzschild--de Sitter spacetime for pure Bell type and isotropic initial states. Using the mode transformations at both horizons, we examined the geometric quantum discord of the accessible modes and the three combinations containing at least one inaccessible mode. The results show that the thermal effects of both horizons, the initial state and the retained modes jointly determine the behavior of geometric quantum discord.

For the accessible modes, increasing the black hole mass $M$ at fixed cosmological constant $\Lambda$ lowers both horizon temperatures and allows the geometric quantum discord to recover. Higher dimensionless frequencies favor its preservation. Different initial states with equal initial geometric quantum discord can exhibit different responses to the horizon effects. In particular, Bell type initial states with complementary parameters $p$ and $\sqrt{1-p^2}$ yield different results over part of the parameter range, reflecting the role of the initial excitation distribution in their response to the thermal effects of both horizons. For the isotropic initial states considered, increasing $q$ enhances the accessible geometric quantum discord.

The inaccessible geometric quantum discord depends on the retained modes. In the parameter range examined, increasing the dimensionless frequency lowers the geometric quantum discord between the inaccessible mode beyond the cosmological horizon and either the accessible black hole side mode or its inaccessible partner, whereas the frequency ordering of the curves for the inaccessible mode inside the black hole and the accessible mode near the cosmological horizon reverses as the black hole mass increases. This reversal reflects competition between the reduced thermal disturbance of the accessible mode near the cosmological horizon and the declining thermal contribution of the inaccessible mode inside the black hole. For the Bell type initial state examined in the low frequency comparison, the geometric quantum discord of all three inaccessible mode combinations approaches a common fixed value as the dimensionless frequency tends to zero at fixed nondegenerate geometry. Furthermore, for the selected parameters with equal initial geometric quantum discord, Bell type and isotropic initial states yield the same geometric quantum discord when the inaccessible black hole mode is paired with either cosmological mode, but differ over part of the range when the accessible black hole side mode is paired with the inaccessible mode beyond the cosmological horizon. The influence of the initial state therefore depends on which modes are retained.

At fixed positive dimensionless frequency, the accessible geometric quantum discord recovers its initial value in the Nariai limit, while all three inaccessible geometric quantum discords vanish. These results reveal how the temperatures of both horizons affect geometric quantum discord in different modes and show that states with equal initial geometric quantum discord can respond differently to the horizon thermal effects.

\section*{Acknowledgments}
This work is supported by the Natural Science Foundation of Hainan Province under Grant No.~125RC744;  the China Scholarship Council (CSC); the Hainan Province Flexible Talent Introduction Collaborative Innovation Center (Yu Changbin); the specific research fund of The Innovation Platform for Academicians of Hainan Province No: YSPTZX202407.


\begin{thebibliography}{99}
\bibitem{OllivierZurek2001}
H. Ollivier and W. H. Zurek, Quantum discord: A measure of the quantumness of correlations, Phys. Rev. Lett. {\bf 88}, 017901 (2001).
\bibitem{HendersonVedral2001}
L. Henderson and V. Vedral, Classical, quantum and total correlations, J. Phys. A: Math. Gen. {\bf 34}, 6899 (2001).
\bibitem{Modi2012}
K. Modi, A. Brodutch, H. Cable, T. Paterek, and V. Vedral, The classical-quantum boundary for correlations: Discord and related measures, Rev. Mod. Phys. {\bf 84}, 1655 (2012).
\bibitem{Adesso2016}
G. Adesso, T. R. Bromley, and M. Cianciaruso, Measures and applications of quantum correlations, J. Phys. A: Math. Theor. {\bf 49}, 473001 (2016).
\bibitem{Datta2008}
A. Datta, A. Shaji, and C. M. Caves, Quantum discord and the power of one qubit, Phys. Rev. Lett. {\bf 100}, 050502 (2008).
\bibitem{Dakic2012}
B. Daki\'c, Y. O. Lipp, X. Ma, M. Ringbauer, S. Kropatschek, S. Barz, T. Paterek, V. Vedral, A. Zeilinger, \v{C}. Brukner, and P. Walther, Quantum discord as resource for remote state preparation, Nat. Phys. {\bf 8}, 666 (2012).
\bibitem{Girolami2014}
D. Girolami, A. M. Souza, V. Giovannetti, T. Tufarelli, J. G. Filgueiras, R. S. Sarthour, D. O. Soares-Pinto, I. S. Oliveira, and G. Adesso, Quantum discord determines the interferometric power of quantum states, Phys. Rev. Lett. {\bf 112}, 210401 (2014).
\bibitem{Dakic2010}
B. Daki\'c, V. Vedral, and \v{C}. Brukner, Necessary and sufficient condition for nonzero quantum discord, Phys. Rev. Lett. {\bf 105}, 190502 (2010).
\bibitem{LuoFu2010}
S. Luo and S. Fu, Geometric measure of quantum discord, Phys. Rev. A {\bf 82}, 034302 (2010).
\bibitem{Radhakrishnan2020}
C. Radhakrishnan, M. Lauri\`ere, and T. Byrnes, Multipartite generalization of quantum discord, Phys. Rev. Lett. {\bf 124}, 110401 (2020).
\bibitem{LiFei2021}
B. Li, C. L. Zhu, X. B. Liang, B. L. Ye, and S. M. Fei, Quantum discord for multiqubit systems, Phys. Rev. A {\bf 104}, 012428 (2021).
\bibitem{Unruh1976}
W. G. Unruh, Notes on black-hole evaporation, Phys. Rev. D {\bf 14}, 870 (1976).
\bibitem{Hawking1975}
S. W. Hawking, Particle creation by black holes, Commun. Math. Phys. {\bf 43}, 199 (1975).
\bibitem{Crispino2008}
L. C. B. Crispino, A. Higuchi, and G. E. A. Matsas, The Unruh effect and its applications, Rev. Mod. Phys. {\bf 80}, 787 (2008).
\bibitem{Fuentes2005}
I. Fuentes-Schuller and R. B. Mann, Alice falls into a black hole: Entanglement in noninertial frames, Phys. Rev. Lett. {\bf 95}, 120404 (2005).
\bibitem{Alsing2006}
P. M. Alsing, I. Fuentes-Schuller, R. B. Mann, and T. E. Tessier, Entanglement of Dirac fields in noninertial frames, Phys. Rev. A {\bf 74}, 032326 (2006).
\bibitem{Martin2010}
E. Mart\'in-Mart\'inez, L. J. Garay, and J. Le\'on, Unveiling quantum entanglement degradation near a Schwarzschild black hole, Phys. Rev. D {\bf 82}, 064006 (2010).
\bibitem{Bruschi2010}
D. E. Bruschi, J. Louko, E. Mart\'in-Mart\'inez, A. Dragan, and I. Fuentes, Unruh effect in quantum information beyond the single-mode approximation, Phys. Rev. A {\bf 82}, 042332 (2010).
\bibitem{Datta2009}
A. Datta, Quantum discord between relatively accelerated observers, Phys. Rev. A {\bf 80}, 052304 (2009).
\bibitem{Brown2012}
E. G. Brown, K. Cormier, E. Mart\'in-Mart\'inez, and R. B. Mann, Vanishing geometric discord in noninertial frames, Phys. Rev. A {\bf 86}, 032108 (2012).
\bibitem{WangDiscord2010}
J. Wang, J. Deng, and J. Jing, Classical correlation and quantum discord sharing of Dirac fields in noninertial frames, Phys. Rev. A {\bf 81}, 052120 (2010).
\bibitem{Wang2014}
J. Wang, J. Jing, and H. Fan, Quantum discord and measurement-induced disturbance in the background of dilaton black holes, Phys. Rev. D {\bf 90}, 025032 (2014).
\bibitem{Haddadi2024}
S. Haddadi, M. A. Yurischev, M. Y. Abd-Rabbou, M. Azizi, M. R. Pourkarimi, and M. Ghominejad, Quantumness near a Schwarzschild black hole, Eur. Phys. J. C {\bf 84}, 42 (2024).
\bibitem{Xiong2026}
Y. Xiong, Z. Pi, T. Zhang, and X. Huang, Quantum discord of mixed states under noisy channels in the curved spacetime, Eur. Phys. J. C {\bf 86}, 317 (2026).
\bibitem{LiHuang2026}
S. Li and X. Huang, Geometric quantum discord in the black hole quantum atmosphere, Eur. Phys. J. C {\bf 86}, 1107 (2026).
\bibitem{Wang2010}
J. Wang, Q. Pan, and J. Jing, Entanglement redistribution in the Schwarzschild spacetime, Phys. Lett. B {\bf 692}, 202 (2010).
\bibitem{Harikrishnan2022}
S. Harikrishnan, S. Jambulingam, P. P. Rohde, and C. Radhakrishnan, Accessible and inaccessible quantum coherence in relativistic quantum systems, Phys. Rev. A {\bf 105}, 052403 (2022).
\bibitem{WuSteering2022}
S. M. Wu and H. S. Zeng, Fermionic steering and its monogamy relations in Schwarzschild spacetime, Eur. Phys. J. C {\bf 82}, 716 (2022).
\bibitem{WuAccessible2024}
S. M. Wu, X. W. Teng, J. X. Li, S. H. Li, T. H. Liu, and J. C. Wang, Genuinely accessible and inaccessible entanglement in Schwarzschild black hole, Phys. Lett. B {\bf 848}, 138334 (2024).
\bibitem{WuTeleportation2023}
S. M. Wu, X. W. Fan, R. D. Wang, H. Y. Wu, X. L. Huang, and H. S. Zeng, Does Hawking effect always degrade fidelity of quantum teleportation in Schwarzschild spacetime?, J. High Energy Phys. {\bf 11}, 232 (2023).
\bibitem{WuLi2025}
S. M. Wu and S. H. Li, Do maximally entangled states always have an advantage over non-maximally entangled states in Schwarzschild black hole?, Eur. Phys. J. C {\bf 85}, 614 (2025).
\bibitem{LiShangWu2025}
S. H. Li, S. H. Shang, and S. M. Wu, Quantum steering for different types of Bell-like states in gravitational background, Phys. Lett. B {\bf 870}, 139895 (2025).
\bibitem{LiAcceleration2025}
S. H. Li, S. H. Shang, and S. M. Wu, Does acceleration always degrade quantum entanglement for tetrapartite Unruh-DeWitt detectors?, J. High Energy Phys. {\bf 05}, 214 (2025).
\bibitem{LiFreeze2026}
S. H. Li, H. C. Yang, R. Y. Xu, and S. M. Wu, Does relativistic motion really freeze initially maximal entanglement?, J. High Energy Phys. {\bf 03}, 218 (2026).
\bibitem{GibbonsHawking1977}
G. W. Gibbons and S. W. Hawking, Cosmological event horizons, thermodynamics, and particle creation, Phys. Rev. D {\bf 15}, 2738 (1977).
\bibitem{BhattacharyaChakrabortty2020}
S. Bhattacharya, S. Chakrabortty, and S. Goyal, Dirac fermion, cosmological event horizons, and quantum entanglement, Phys. Rev. D {\bf 101}, 085016 (2020).
\bibitem{BhattacharyaGaur2020}
S. Bhattacharya, H. Gaur, and N. Joshi, Some measures for fermionic entanglement in the cosmological de Sitter spacetime, Phys. Rev. D {\bf 102}, 045017 (2020).
\bibitem{WuCoherence2023}
S. M. Wu, C. X. Wang, D. D. Liu, X. L. Huang, and H. S. Zeng, Would quantum coherence be increased by curvature effect in de Sitter space?, J. High Energy Phys. {\bf 02}, 115 (2023).
\bibitem{Jiang2025}
X. Y. Jiang, X. L. Huang, and S. M. Wu, Cosmological entanglement of initial multipartite states, Eur. Phys. J. C {\bf 85}, 851 (2025).
\bibitem{Bhattacharya2018}
S. Bhattacharya, Particle creation by de Sitter black holes revisited, Phys. Rev. D {\bf 98}, 125013 (2018).
\bibitem{QiuTraschen2020}
Y. Qiu and J. Traschen, Black hole and cosmological particle production in Schwarzschild de Sitter, Class. Quantum Grav. {\bf 37}, 135012 (2020).
\bibitem{BhattacharyaJoshi2022}
S. Bhattacharya and N. Joshi, Entanglement degradation in multi-event horizon spacetimes, Phys. Rev. D {\bf 105}, 065007 (2022).
\bibitem{Liu2023}
Q. Liu, S. M. Wu, C. Wen, and J. Wang, Quantum properties of fermionic fields in multi-event horizon spacetime, Sci. China Phys. Mech. Astron. {\bf 66}, 120413 (2023).
\bibitem{Wu2024}
S. M. Wu, J. X. Li, X. W. Fan, W. M. Li, X. L. Huang, and H. S. Zeng, Gaussian quantum steering in multi-event horizon spacetime, Eur. Phys. J. C {\bf 84}, 176 (2024).
\bibitem{WuFermionic2025}
S. M. Wu, J. X. Li, Y. X. Wang, S. H. Shang, and J. Lu, Fermionic steering in multi-event horizon spacetime, Eur. Phys. J. C {\bf 85}, 54 (2025).
\bibitem{refmi2025}
G. W. Mi, X. Huang, S. M. Fei, and T. Zhang, Genuine four-partite Bell nonlocality in the curved spacetime, Eur. Phys. J. C {\bf 85}, 354 (2025).
\bibitem{Major2022}
S. A. Major, Energy for accelerating observers in black hole spacetimes, Phys. Rev. D {\bf 105}, 104050 (2022).
\bibitem{Gomberoff2003}
A. Gomberoff and C. Teitelboim, De Sitter black holes with either of the two horizons as a boundary, Phys. Rev. D {\bf 67}, 104024 (2003).
\bibitem{Sekiwa2006}
Y. Sekiwa, Thermodynamics of de Sitter black holes: Thermal cosmological constant, Phys. Rev. D {\bf 73}, 084009 (2006).
\bibitem{WangBattery2026}
X. Wang, X. Huang, Z. Ma, S. M. Fei, and T. Zhang, Noise is not always detrimental: The capacity of quantum batteries is enhanced in black holes, arXiv:2604.05325 (2026).
\bibitem{Horodecki1999}
M. Horodecki and P. Horodecki, Reduction criterion of separability and limits for a class of distillation protocols, Phys. Rev. A {\bf 59}, 4206 (1999).
\end{thebibliography}
\end{document}